

\input harvmac.tex

\noblackbox

\def\sign{{\rm sign}}
\def\sech{{\rm sech}}
\def\half{{\scriptstyle{1\over2}}}

\def\e{\epsilon}

\nref\hofref{M.Ya.Azbel', Zh.Eksp.Teor.Fiz. {\bf 46}, (1964) 939 [Sov.Phys.
JETP {\bf 19}, (1964) 634] \semi
D.R. Hofstadter, Phys. Rev. {\bf B14} (1976) 2239 \semi
G. H. Wannier, Phys. Status Solidi {\bf B88} (1978) 757.}
\nref\static{The option of static randomness, in which
a fixed potential is chosen from a random ensemble, has been recently studied
by
Azbel.}
\nref\dissqm{A.~O.~Caldeira and A.~J.~Leggett, Physica  {\bf 121A}(1983) 587;
Phys. Rev. Lett. {\bf 46} (1981) 211; Ann. of Phys. {\bf 149} (1983) 374.}
\nref\thetar{J.~L.~Cardy, ``Duality and the Theta Parameter in Abelian
Lattice Models'', Nucl. Phys. {\bf B205} 17 (1982).}
\nref\osdqm{C.G. Callan, L. Thorlacius, ``Open String Theory as Dissipative
Quantum Mechanics'', Nucl. Phys. {\bf B329} (1990) 117.}
\nref\dyson{The Dyson chain with $1/r^n$ interactions for $n\le 2$ is
known to have phase transitions.  See references \cardy,
\kjaer, and references therein.}
\nref\fisher{M.~P.~A.~Fisher and W.~Zwerger, Phys. Rev. {\bf B32} (1985) 6190.}
\nref\ghm{F.~Guinea, V.~Hakim and A.~Muramatsu, Phys. Rev. Lett.
{\bf 54} (1985) 263.}
\nref\kleb{I. Klebanov, ``String Theory in Two Dimensions", Princeton preprint
PUPT-1271, July 1991.}
\nref\schmid{A.~Schmid, Phys. Rev. Lett. {\bf 51} (1983) 1506.}
\nref\defjah{D.~Freed and J.~A.~Harvey,``Instantons and the Spectrum of
Bloch Electrons in a Magnetic Field'', Phys. Rev. {\bf B41} (1990) 11328.}
\nref\kjaer{K. Kjaer and H. Hillhorst, ``The Discrete Gaussian Chain
with $1/r^n$ Interactions: Exact Results'',
Journal of Statistical Physics {\bf 28} (1982) 621.}
\nref\cardy{J.~L.~Cardy, ``One-dimensional Models with $1/r^2$
Interactions'', J. Phys. {\bf A14} (1981) 1407.}
\nref\frohlich{J.~Fr\"ohlich and B.~Zegarlinski, ``The Phase Transition
in the Discrete Gaussian Chain with $1/r^2$ Interaction Energy'',
ETH-TH/90-25.}
\nref\bulgy{S.~A.~Bulgadaev, JETP Lett. {\bf 39} (1984) 315 .}

\Title{\vbox{\baselineskip12pt\hbox{PUPT-1260}}}
{Phase Diagram of the Dissipative Hofstadter Model}

\baselineskip=11.5pt
\bigskip
\centerline{Curtis G. Callan, Jr. and Denise Freed
\footnote{*}{Now at: Center for Theoretical Physics, Massachusetts
Institute of Technology, Cambridge MA 02138.}}
\centerline{\it Department of Physics, Princeton University}
\centerline{\it Princeton, NJ 08544}
\centerline{\it Internet: callan@pupphy.princeton.edu,
freed@mitlns.mit.edu}

\bigskip
\centerline{\bf Abstract}
A quantum particle moving in a uniform magnetic field and periodic potential
(the Hofstadter model) has an energy band structure which varies in a
discontinuous fashion as a function of the magnetic flux per lattice unit cell.
In a real system, randomness of various kinds should ``smooth out''
this behavior in some way. To explore how this happens,
we have studied the dissipative quantum mechanics of the Hofstadter model.
We find, by virtue of a duality in a two-dimensional space parametrized
by the dissipation constant and the magnetic field strength, that
there are an infinite number of phase transition lines, whose density grows
without limit as the dissipation goes to zero and the model reduces to the
original Hofstadter model. The measurable quantity of
greatest interest, the mobility, can be determined exactly in most of parameter
space. The critical theory on the phase transition lines has yet to be
characterized in any detail, but it has reparametrization invariance
and defines a set of nontrivial backgrounds for open string theory.

\Date{October 1991}
\eject

\newsec{Hofstadter Model and Dissipation: Introduction and Issues}

The quantum mechanics of an electron moving in two dimensions subject to a
uniform magnetic field and a periodic potential (which we will call the
Hofstadter problem for short)
has a remarkable fractal energy band structure \hofref\ which varies
discontinuously as a function of the number of flux quanta per
lattice unit cell. Such behavior must, however, be ``smoothed out'' by the
inevitable randomness of a real physical system. An interesting model of
dynamic randomness \static\ (due ultimately to weak coupling of the electron
to lattice vibrations) is provided by the dissipative quantum mechanics of
Caldeira and Leggett (DQM) \dissqm\ and the topic of this paper is the
application of DQM to the Hofstadter problem. A key point is that
DQM converts quantum mechanics into one-dimensional statistical mechanics
(the one dimension being Euclidean time on the electron's world line) with
long-range interactions strong enough to support phase transitions between,
as it turns out, localized and delocalized long-time behavior.
Thus, the discontinuities of the original
quantum mechanics are replaced by phase transitions and the problem is to
map out the phase diagram (in the magnetic field-dissipation constant plane).
Presumably the density of phase transitions grows as dissipation decreases,
recreating discontinuous dependence on the magnetic field in the
zero-dissipation limit. We are able to show, using a duality argument, that
this is exactly what happens and that the phase diagram is what was found
years ago \thetar\ in the study of theta-dependence of the
confinement-deconfinement transition in Abelian
lattice gauge theories! Away from the critical lines, the behavior of the
system for most of the parameter space is very simple: the long-time
mobility, for example, is given by the trivial theory for some value of the
magnetic field and no potential. The critical theories, however, are definitely
non-trivial and we have only been able to extract limited information about
them. Apart from their condensed matter interest, they represent new solutions
of open string theory \osdqm . Our results at least indicate that there is a
rich vein of one-dimensional critical theories to explore, once the
appropriate tools have been developed.

The outline of this paper is as follows.  Section~2 contains a review of the
dissipative Hofstadter model.  In Section~3 we sketch the arguments showing
that, at specific critical points, the theory is at a renormalization group
fixed point and the mobility can be calculated exactly.  In the fourth section,
we derive the approximate duality transformations of the dissipative Hofstadter
model, and, in the following section, we demonstrate that the duality
transformations are exact for a discrete version of the model.  In section~6,
we use these transformations to map out the phase diagram of the dissipative
Hofstadter model.  In the final section, we present a brief discussion of our
results.

\newsec{Setup of the Dissipative Hofstadter Model}

We begin with a brief outline of dissipative quantum mechanics.
For details the reader is referred to
\dissqm\ and \osdqm\ . The quantum mechanics of a particle subject to a
scalar potential $V$ and vector potential $\vec A$ is described by the
Euclidean path integral
\eqn\eucpi{Z_{QM} = \int [{\cal D}X(t)] \, e^{-S_{QM}[X]}}
with
\eqn\qmact{S_{QM}[X] = \int dt \bigl\{{\half}M\dot{\vec X}^2 + V(X)\bigr\}
                           +i\int dt\, A_i (X)\,\dot X^i \, .}
If the quantum variable, $X^i$, is in some sense macroscopic (a good
example is the trapped magnetic flux in a Josephson junction) there will
typically exist an infinite set of degrees of freedom to which $X^i$ is at
least weakly coupled and which give rise to dissipative effects on the motion
of $X^i$. In the classical limit, dissipation can be described by
adding a phenomenological friction term $-\eta \dot X^i$ to the equation
of motion. It is then natural to ask whether there is a correspondingly simple
and universal way to express the {\it quantum\/} effects of weak
dissipation.  A practical motivation is to assess whether coherent quantum
effects, such as tunneling, can really be observed in macroscopic systems.

Caldeira and Leggett \dissqm\ addressed this problem via a simple model:
In addition to the coordinates $X^i$, let there be a bath of harmonic
oscillators, $q_\alpha$, with a distribution of frequencies, $\omega_\alpha$,
coupled linearly to the $X^i$ with weak coupling strengths, $C_\alpha$.
If the parameters satisfy the functional condition
\eqn\parcon{\sum_\alpha {C_\alpha^2 \over 2\omega_\alpha}\;
           \delta (\omega {-}\omega_\alpha ) = {\eta\,\omega \over \pi}~,}
then, when the oscillators are `integrated out' of the classical equations of
motion for $X^i$, they supply the canonical $-\eta \dot X^i$ friction term.
Since the dependence of the action on the $q_\alpha$ is only quadratic,
they can be explicitly integrated out of the quantum path integral also.
The result is a new path integral over the $X^i$ alone, where the quantum
effect of friction is contained in a non-local term whose strength is set by
the classical friction constant $\eta$:
\eqn\disspi{Z_{QM}^\eta =\int [{\cal D}X(t)] \, e^{-S_{QM}[X] - S_\eta [X]}}
where
\eqn\disske{S_\eta [X]={\eta\over 4\pi\hbar}\int\limits_{-\infty}^\infty
 dt\,dt' \; {\bigl(\vec X(t){-}\vec X(t')\bigr)^2 \over (t{-}t')^2}\,.}
Because of the non-locality of the $\eta$-term, the path integral \disspi\
is effectively that of a one-dimensional statistical system with long-range
interactions. Such systems, unlike one-dimensional {\it local} systems, have
phase transitions (the classic example being the Ising chain with $1/r^2$
interactions \dyson\ ). In the DQM context, the phase transitions are
between different regimes of long-time behavior of Green's functions (typically
between localized and delocalized). The crucial qualitative information about
DQM therefore concerns the phase structure of its long-time behavior and
we will analyze the dissipative Hofstadter model from that point of view.

To study the Hofstadter model, we specialize \qmact\ as follows:
$\vec X=(x,y)$ is taken to be two-dimensional, the potential is taken to be
periodic with period $a$ in both directions, and of strength $V_0$, so that
\eqn\perpot{V(x,y)= V_0 \cos(2\pi x/a) + V_0 \cos(2\pi y/a)~;}
and the magnetic field is taken to be uniform and in a linear gauge:
\eqn\magfld{(A_x,A_y)=\half (By,-Bx)~.}
Leaving aside the potential term, the action \qmact\ is Gaussian:
\eqn\hofgauss{S_{g}={\eta\over 4\pi\hbar}
	\int_{-T/2}^{T/2}dtdt^\prime{(\vec X(t)-\vec X(t^\prime))^2
		\over (t-t^\prime)^2}
		+{M\over 2\hbar}\int dt \dot {\vec X}^2
			+{ieB\over 2\hbar c}\int dt (\dot x y-\dot y x)~.}
The generating functional for for the full theory is then given by
\eqn\disspigf{Z[\eta,B,V,F]=\int\left[DX(t)\right]e^{-S_g[X]
                            -{1\over\hbar}\int V(\vec X) dt
                            -S_F[X]},}
where $S_F[X]$ is a linear source term,
\eqn\SFdef{S_F[X]=\int\vec F(t)\cdot\vec X(t) dt.}
In order to determine the phase structure of the dissipative Hofstadter model,
we will be concentrating on the properties of the partition
function and two-point function, which are obtained from $Z[\eta,B,V,F]$
as follows:
\eqn\disspf{Z[\eta,B,V]=Z[\eta,B,V,0]}
and
\eqn\tpdef{\langle X^\mu(t_1) X^\nu(t_2) \rangle(\eta,B,V) =
         \left.  {1\over Z[\eta,B,V,0]}
        {\delta^2 Z[\eta,B,V,F]\over\delta F^\nu(t_1) \delta
F^\mu(t_2)}
        \right|_{F=0}.}

It is useful to define the dimensionless parameters
$\alpha= \eta a^2 / 2\pi\hbar$ and $\beta= eBa^2 / 2\pi\hbar c$,
to rescale $\vec X$ by $1/a$,
and to reexpress the Gaussian action in Fourier space:
\eqn\fourgaus{S_g=\half\int{d\omega\over 2\pi} \{
	({\alpha\over 2\pi}|\omega|+{Ma^2\over 2\hbar}\omega^2)
	\delta_{ij}+2\pi\beta\e_{ij}\omega \}~
		 \tilde X^*_i(\omega)\tilde X_j(\omega)~.}
One sees that the dissipation and magnetic field terms are dimension one
operators (they grow like the first power of $\omega$) and therefore marginal
from the renormalization group point of view. On the other hand, the ordinary
kinetic energy term is dimension two and therefore irrelevant: it just acts
as a regulator for short times and has no effect on universal critical
properties. We will set it to zero from now on and use a more convenient
short-distance cutoff wherever needed.

The action \fourgaus\ with $M=0$ plays a very important role: it is the
Gaussian fixed point theory which governs the long-time behavior of the
system whenever the potential term \perpot\ is irrelevant. We will shortly
identify the region in the $\alpha$-$\beta$ plane where this condition is met.
Let us first summarize the essentials of the Gaussian theory: all information
is contained in the coordinate two-point function
$\langle X_i(t)X_j(t^\prime)\rangle$, the inverse of the quadratic form in
\fourgaus\ . We will call it $D_{ij}(t-t^\prime;z)$, using the complex number
$z=\alpha+i\beta$ to summarize the parameters on which it depends. Explicitly,
\eqn\twopt{D_{ij}(t-t^\prime;z)=-{\alpha\over\alpha^2+\beta^2}
	\log(t-t^\prime)^2~\delta_{ij}-i{\pi\beta\over\alpha^2+\beta^2}
		\sign(t-t^\prime)\e_{ij}~.}
The logarithmic growth of $D(t)$ clearly indicates that the Gaussian system
is on the borderline between localization and delocalization. The coefficient
of the $\delta_{ij}$ term is called, for obvious reasons, the mobility ($\mu$).
For zero magnetic field, $\mu\to\infty$ as dissipation goes to zero,
while for finite field, the mobility goes to zero in the
same limit since the particles become stuck in Landau levels. The coefficient
of the $\e_{ij}$ term is essentially a Hall coefficient: it measures the
response transverse to an applied field.

In Fourier space, the two-point function is
\eqn\fstwopt{\tilde D_{ij}(\omega;z)
             ={1\over |\omega|} M_{ij}(\omega;{1\over z}),}
where
\eqn\Mfsdef{M_{ij}(\omega,x+iy)
   = x\delta_{ij}+y{|\omega|\over\omega} \epsilon_{ij}.}
In Section~6, we will make use of the following simple rules for the addition
and multiplication of $M_{ij}$:
\eqn\Madd{M_{ij}(\omega;z_1)+M_{ij}(\omega;z_2) = M_{ij}(\omega;z_1+z_2);}
\eqn\Mmult{M_{ij}(\omega;z_1)M_{ij}(\omega;z_2)=M_{ij}(\omega;z_1 z_2);}
and
\eqn\Mcmult{cM_{ij}(\omega;z_1)=M_{ij}(\omega;c z_1)}
for all $z_1,z_2\in C$ and $c\in R$.

Given the explicit Gaussian two-point function, one can easily calculate the
effective dimensionality of a perturbation such as \perpot\ and identify the
regions in the $z$-plane where it is irrelevant.
This is a simple one-loop renormalization group calculation and it is easy to
show that \perpot\ is marginal when the mobility coefficient
${\alpha / (\alpha^2+\beta^2)}$ is unity and irrelevant when
$\alpha/(\alpha^2+\beta^2)$ is greater than one. This condition
defines a circle in the
$z$-plane, of radius one-half and centered at $z=(\half,0)$, inside of which
\perpot\ is irrelevant and on which it is marginal. For zero field, the
critical value is $\alpha=1$, a fact which has been known for a long time
\fisher\ . Outside the circle, \perpot\ is relevant and, if one knew that
there were only two phases, one would have to identify that entire region with
a localized phase. This is almost certainly correct on the zero magnetic
field line, but, given
the rich dual and self-similar structure of the pure quantum mechanics of
the Hofstadter problem, certainly wrong for non-zero field.

Two loosely related problems arise at this point. The first is to map out the
phase structure in the region outside the ``first circle'', where the periodic
potential is relevant. We believe we have a complete solution of this problem.
As will be explained in subsequent sections, the key to the solution is an
approximate
duality symmetry (generated by two transformations $z\to 1/z$ and $z\to z+i$)
which relates behavior in different regions of the $z$-plane.
The second problem is to characterize the family of
critical theories which live on the boundary of the first circle (and its
images under duality). At present we know very little about them, although,
as we shall explain in the next section, they become particularly simple at the
points where $\beta/\alpha$ is integer.  There is a further motivation
for studying these critical theories, whose elaboration must
await another paper. It is that, as explained in \osdqm\ , these critical
theories satisfy an infinite set of Ward identities deriving from
reparametrization invariance, possess a manifest $SL(2,R)$ symmetry, and
represent nontrivial solutions of open string theory!

\newsec{Fermionization and Exact Fixed Points}

In this section, we summarize some of the properties of the critical theories
at the points on the ``first circle'' with integral $\beta/\alpha$.
Using a convenient regulator, we show that at these points the theory has no
logarithmic divergences
and is therefore at a renormalization group fixed point.
In addition, we show that the theory can be fermionized and that
the free energy and mobility can be exactly calculated.

In order to calculate the partition function and correlation functions
of the $\dot X^\mu$'s,
it is most convenient to expand the path integral in powers of the cosine
potential. We then have to evaluate correlators of products
of $\dot X^\mu$'s and $e^{\pm iX^\nu}$'s (the latter coming from expanding the
cosines as a sum of exponentials) and the result can
be written in terms of integrals over products of factors of the type
${\langle \dot X^\mu(t_i) X^\nu(t_j)\rangle}_0$ and
$e^{{\pm\langle X^\mu(t_i) X^\nu(t_j)\rangle}_0}$, where the expectation value
is with respect to the gaussian propagator $D_{ij}(t-t';z)$ and the integrals
are over the $t$ coordinates of the potential insertions. In order to get
finite results at all stages of the calculation, we need both an infrared and
an
ultraviolet cutoff. We obtain the former by putting the parameter $t$
on a circle of circumference $T$, and the latter by
multiplying the Fourier-transformed propagator $\tilde D_{ij}(\omega;z)$
by $e^{-\epsilon {|\omega|}}$ (we do this {\it instead} of giving the particle
an explicit mass). With these regulators \twopt\ is no longer correct, but
it is not too hard to show that
\eqn\regxdotx{\eqalign {{\langle \dot X^\mu(t_1)X^\nu(t_2)\rangle}_0=&
               -\left({2\pi i\over T}\right){\alpha\over \alpha^2 + \beta^2}
    \left[{w_1^2-w_2^2\over(w_1-e^{-\epsilon}w_2)(w_1-e^\epsilon w_2)}\right]
		\delta^{\mu\nu} \cr
   &+ \quad\left({2\pi i\over T}\right){\beta\over \alpha^2 + \beta^2}
     \left[1+{(e^\epsilon-e^{-\epsilon})w_1 w_2\over
      (w_1-e^{-\epsilon}w_2)(w_1-e^\epsilon w_2})\right]\epsilon^{\mu\nu},}}
\eqn\regexex{\eqalign{
e^{\pm{\langle X^\mu(t_1) X^\nu(t_2)\rangle}_0} =&
       \left[-{e^\epsilon w_1 w_2\over
      (w_1-e^\epsilon w_2)(w_1-e^{-\epsilon} w_2)}\right]^
     {\pm {\alpha\over \alpha^2 + \beta^2}\delta^{\mu\nu}}\cr
&\qquad \times\left[{w_1-w_2 e^\epsilon\over w_2-w_1 e^\epsilon}\cdot
 {w_1\over w_2} \right]^{\pm\epsilon^{\mu\nu}{\beta\over\alpha^2+\beta^2}}
  .}}
where we have defined a new position coordinate $w_j=e^{2\pi it_j/T}$ which
lies on the unit circle. Under this transformation,
integrals over $t_j$ are converted into contour integrals of
$w_j$ around the unit circle according to
$\int dt \to {T\over 2\pi i}\oint{dw_j\over w_j}$.
We note that whenever $\alpha/(\alpha^2 + \beta^2) = 1$ and $\beta/\alpha$ is
an integer, the expressions in equations \regxdotx\ and \regexex\ are
rational functions of $w_j$ and $e^\epsilon$, so the integrands for the
partition function and n-point functions will also be rational functions.
Furthermore, because the integrand factors into a product of simple poles, we
can perform the integrals by successively extracting the residues of the
poles in the integration variables which lie within the unit circle.
At each stage we are left with a
rational function of the remaining $w_j$'s and $e^\epsilon$.  This implies
that, in the end, we can only get pole divergences in $\epsilon$ as
$\epsilon\to0$, and no logarithmic divergences.  In particular, for correlators
of marginal operators such as $\dot {\vec X}(t)$, which can at worst have
logarithmic divergences, there can be no divergences at all.
As a result, when $\alpha/(\alpha^2 + \beta^2) =1$ and
${\beta /\alpha} \in Z$, the theory should be at a zero of the
$\beta$-function.  As we will show elsewhere, this also implies that the
theory satisfies an infinite set of Ward Identities, which
derive from reparametrization invariance. (As was explained in \osdqm , these
Ward identities are what make possible a string theory interpretation of
systems of this kind.) This is very much like
the $2D$ case, where scale invariance implies conformal invariance.
We note that the absence of logarithmic divergences occurs for
any value of the potential, $V_0$, which means that these fixed-point theories
are really fixed lines.

We can use the regulator and contour integration procedure just described
to calculate the partition function and $n$-point functions to any order
in $V_0$, but it becomes extremely tedious for higher orders. At the
special point $\alpha=1$, $\beta=0$, we can obtain a number of results
exact to all orders in $V_0$ by fermionizing the unregulated theory and then
regulating the fermionic theory. We observe that if we express the $X$'s
in terms of two free fermions $\psi_+(t)$ and $\psi_-(t)$ according to
\eqn\eixferm{e^{iX(t)}=\psi_+^\dagger(t)\psi_-(t),}
and
\eqn\xdotferm{\dot X(t) = {2\pi i\over T}\left[\psi_+^\dagger(t)\psi_+(t)-
                \psi_-^\dagger(t)\psi_-(t)\right].}
and define the fermion  propagators to be
\eqn\ppprop{\langle \psi_+^\dagger(t_1) \psi_+(t_2)\rangle =
           \langle \psi_-^\dagger(t_1) \psi_-(t_2)\rangle =
           {i\sqrt{w_1 w_2} \over w_2-w_1}}
and
\eqn\papb{\langle \psi_+^\dagger(t_1) \psi_-(t_2)\rangle =
         \langle \psi_-^\dagger(t_1) \psi_+(t_2)\rangle = 0~,}
we reproduce all the bosonic correlators of $\dot X$'s and $e^{\pm iX}$.
Note that the set of operators under discussion here generates an $SU(2)$
algebra and is a particular example of the discrete states which have been
extensively discussed recently in two-dimensional quantum gravity \kleb .

Since the potential term, and all the operators of interest, are now
quadratic in free fermions, the unregulated problem reduces to free field
theory. The ultraviolet-regulated theory might not, however, and care must
be taken here. We regulate by multiplying the Fourier transform of the fermi
propagators by $e^{-\epsilon|\omega|}$, and again find that all the integrals
for the free energy and $n$-point functions are contour integrals of factored
rational polynomials. This time, all the integrands are just products of
loops of the form $\prod_{i=1}^l 1/(w_i-e^{-\epsilon}w_{i+1})$,
with $w_{l+1}=w_1$, so we can easily perform the integrals at any order in
$V_0$. Furthermore, at least for the partition function and the two-point
function, it is easy to resum the expansion in powers of the potential.
The result for the free energy density is
\eqn\freeen{F = {1\over 2\epsilon} \sum_{N=1}^\infty {1\over N^2}(-1)^N
                 \left({V_0T\over2} e^{-\langle X(0) X(0)\rangle}\right)^{(2N)}
                  + O(\epsilon),}
while the result for the two point function is
\eqn\toopt{\langle\dot X(t_1) \dot X(t_2) \rangle =
	\left({2\pi\over T}\right)^2
         \left({1\over 1+ ({V_0T\over2}e^{-\langle X(0) X(0)\rangle})^2}\right)
         {2w_1 w_2\over (w_1-w_2)^2}.}
Note that both of these expressions depend on the peculiar dimensionless
parameter $V_0Te^{-<X(0)X(0)>}$. In the limit that the ratio of infrared to
ultraviolet cutoff is large, this parameter reduces to $V_0/\epsilon=
\tilde V_0$. The only renormalization needed to render the two-point function
(and, indeed, all n-point functions of the $\dot X$'s) finite is to
choose $\tilde V_0$ to be finite. This doesn't quite render the free energy
density finite: it has a $1/\epsilon$ piece which we must subtract away by
hand. We only attach physical significance to the finite piece left
over after this subtraction and we note that in this
particular case ($\alpha = 1$ and $\beta = 0$) this finite piece is zero.
Note also that the functional form of the two-point function follows from
$SU(1,1)$ invariance if $\dot X$ transforms with weight unity. Both of these
points and their string theory significance are explained at greater length
in \osdqm.

Apart from an overall $V_0$-dependent, finite, multiplicative renormalization,
\toopt\ is the same as the Gaussian two point function $D(t-t';1)$ transformed
to the unit circle (consider \regxdotx\ at $\alpha=1, \beta=0$).
So, at the $\alpha=1$ fixed point, the ratio of the renormalized mobility
to its Gaussian value has a specific dependence on the potential strength:
\eqn\mueff{ \mu_{eff} /\mu_{Gauss}=
	(1+(e^{-\langle X(0) X(0) \rangle}V_0T/2)^2)^{-1}
	=(1+(\tilde V_0/2)^2)^{-1}~. }
This is in rough accord with the results of Guinea et.~al. \ghm. Since they
worked in the tight-binding approximation, rather than in the weak-potential
expansion used here, the renormalization of the Gaussian two-point function
is different in the two cases. Rather surprisingly,
even for non-vanishing potential strength,
the higher n-point functions turn out to vanish for non-coincident points
(i.e. when $t_i\ne t_j$ for $1\le i,j\le n$). On the other hand, when all the
$t_i$ are equal, the regulated $2n$-point functions are proportional to
$1/\epsilon^{2n}$. Thus, the $n$-point functions for $n > 2$ are zero
except for contact terms.  Because of the rescaling of the $2$-point
function and the contact terms for higher $n$-point functions, the fixed point
theory when $\alpha=1$ does not appear to be a simple gaussian
theory. Precisely what it is is not yet known in any detail.

It is also possible to fermionize the theory at the other special
points where $\beta/\alpha \in Z$ and $\alpha=\alpha^2+\beta^2$. Although the
$\dot X$'s require more care to fermionize because
$\langle e^{iY(t_1)} \dot X(t_2)\rangle \propto {d\over dt}
\sign(t_1-t_2)$ is ill-defined without the regulator,
the results for the free energy and 2-point functions are very similar to what
we found at the $\alpha=1$ fixed point.

\newsec{Duality Symmetries from Coulomb Gas and Instanton Methods }

We will now show that the dissipative Hofstadter model has at least an
approximate duality symmetry
under the modular group consisting of the transformations
$z\to1/z$, $z\to z+i$ and all their compositions, where $z=\alpha +
i\beta$.  In Section~6 we will use this duality to obtain an
overall picture of the phase structure of the theory.
To demonstrate the symmetry, we begin by deriving a generalized Coulomb gas
representation of the DQM path integral.  The first duality symmetry then
follows from a manifest invariance of the Coulomb gas.  Next, we derive a
second Coulomb gas representation, which is related to the first by a
transformation of $z$ and $V_0$.  Insofar as both Coulomb gases are valid
representations of the same model, we obtain the second duality transformation.
These two transformations on $z$ are enough to generate the full modular group.
This strategy was first applied to DQM {\it without} a
magnetic field by Schmid \schmid\ to obtain restricted duality under the
transformation $\alpha\to 1/\alpha$ of the dissipation constant only.
It was also used in reference \defjah\ for the Hofstadter model with no
dissipation.

For the first derivation of the Coulomb gas, we simply
expand the path integral in powers of the potential and further expand
the cosine potential as a sum of
positive and negative frequency exponentials:  $\cos X(\tau_j)=
\half\sum_{e_j=\pm1}e^{ie_j X(\tau_j)}$.  Because the potential
\perpot\ is a sum of terms periodic in $X$ and $Y$, we must actually
define a two-dimensional vector of charges $\vec e_i$.  A collection of
$n$ of these charges, located at times $\tau_i$, can be associated with
a charge density
\eqn\rhodef{\vec\rho^n(\tau)=\sum_{j=1}^{n} \vec e_j\delta(\tau-\tau_j).}
With these definitions, the generating function \disspigf\ becomes
\eqn\ZCGcalc{Z[z,V_0,F] = \int [D \vec X(t)] \sum_{n=0}^\infty
         \int d\tau_1\dots d\tau_n
    \left({V_0\over 2}\right)^n {1\over n!}
    \sum_{\vec e_j = {(\pm 1,0)\atop(0,\pm1)}}
     e^{-S_q},}
where
\eqn\SqCGdef{S_q
      =S_g+S_F-i\int \vec X(\tau)\cdot\vec \rho^n(\tau)  d\tau.}
Because $S_q$ is quadratic, the $\vec X$ integration can be done exactly,
with the result
\eqn\ZZgen{\eqalign{Z[z,V_0,F]
    =&C Z_{\rm gen}\left[S_{\rm gen}(z,F);{V_0\over 2}\right]\cr
    =&C \sum_N \sum_{\{\vec e_i\}}\int d\tau_1\dots d\tau_{2N}
        {1\over(2N)!}\left({V_0\over2}\right)^{2N}
         \exp[-S_{\rm gen}(z,F)],}}
where the action, $S_{\rm gen}(z,F)$, is given by
\eqn\SCGgen{S_{\rm gen}(z,F) = {1\over2}\int d\tau d\tau'
     \left[i\vec F(\tau)+\vec \rho^N(\tau)\right]^\dagger D(\tau-\tau';z)
     \left[i\vec F(\tau')+\vec \rho^N(\tau')\right].}
The allowed values for the vector ``charges'' $\vec e_i$ in the expression
for $Z_{\rm gen}$ are $(\pm1,0)$ and $(0, \pm1)$.
Integrating over the zero mode of $\vec X$ gives
the additional requirement of charge neutrality:
$\sum_{i=1}^{2N} \vec e_i = 0$.

To obtain the partition function, we set $F=0$ in equation \ZZgen. It will
be useful to note that the resulting object is identical to the grand
canonical partition function for a certain generalized Coulomb gas.
Let us define a Coulomb gas interaction ``energy'' between a collection
of charges $\vec e_i$ at locations $\tau_i$ by
\eqn\SCGdefa{S_{CG}(A+iB)={1\over2}\sum_{i,j}
            \vec e_i\cdot D(\tau_i-\tau_j;{1\over A+iB})\cdot\vec e_j~,}
where $D$ is the Gaussian two-point function of \twopt.
Writing out the full expression for $D$, we obtain the explicit expression
\eqn\fullSCG{S_{CG}(A+iB)=A\sum_{i<j}\vec e_i\cdot\vec e_j \ln(t_i-t_j)^2
     +i\pi B\sum_{i<j}\sign(\tau_i-\tau_j)\epsilon^{\mu\nu}e_i^\mu e_j^\nu.}
This is just a collection of two-body interaction terms, with real and
imaginary parts whose strengths are set by $A$ and $B$ respectively. The
real part looks like a standard Coulomb potential term. The
imaginary part is of course unconventional and that is why we refer to this
as a generalized Coulomb gas. Its partition function is
\eqn\CGpfn{\eqalign{Z_{CG}[A+iB,\zeta]
    =& \sum_N \sum_{\{\vec e_i\}}\int d\tau_1\dots d\tau_{2N}
        {1\over(2N)!}\left({\zeta}\right)^{2N}
         \exp[-S_{CG}(A+iB)],}}
where $\zeta$ is the fugacity and the sum is over even numbers of charges only
because we impose overall charge neutrality. The imaginary part of the
interaction energy gives a phase to each term in the partition sum. In this
regard it is similar to the $\theta$ term of QCD.

The partition function of interest to us, \ZZgen\ evaluated at $F=0$,
can thus be written in terms of a generalized Coulomb gas as follows:
\eqn\ZZCG{Z[z,V_0]=CZ_{CG}[{1/ z}, {V_0/2}].}
In short, the original system is equivalent to
a generalized Coulomb gas in which each particle has charge
$\vec e_j=(\pm1,0)$ or $(0,\pm1)$ and fugacity $\zeta=V_0/2$,
and the bonds between particles are assigned both an energy and a
phase determined by the parameter $z=1/(A+iB)$.

By varying the generating function \ZZgen\ with respect to $\vec F$ and
setting $\vec F=0$ one gets expressions for n-point functions of the $X$'s
in terms of Coulomb gas correlators of the charge density $\vec\rho$.
For the two-point function, in particular, we obtain
\eqn\tpcg{\langle\tilde X^\mu(\omega) \tilde X^\nu(-\omega)\rangle(z,V_0) =
       \tilde D^{\mu\nu}(\omega;z) -
      \tilde D^{\mu\sigma}(\omega;z)\,
   \langle\tilde\rho^\sigma(\omega)
	\tilde\rho^{\lambda}(-\omega)\rangle({1/ z}, V_0/2) \,
      \tilde D^{\lambda\nu}(\omega;z),}
where by
$\langle\tilde\rho^\sigma(\omega)
	\tilde\rho^{\lambda}(-\omega)\rangle(1/z, V_0/2)$
we mean the density-density correlation function in the Coulomb gas with
coupling parameter $1/z$ and fugacity $V_0/2$.
We should emphasize that although we have expanded in powers of $V_0$, no terms
have been dropped, so our results to this point are exact.

The generalized Coulomb gas described by $Z_{CG}[A+iB,\zeta]$ possesses a
symmetry under shifts in $A+iB$.  Because all the charges are
integral, the phase factor appearing in a generic term in the
partition sum,
\eqn\phase{ \exp[i\pi B\sum_{i<j}\sign(\tau_i-\tau_j)
	\epsilon^{\mu\nu} e_i^\mu e_j^\nu]~,}
is invariant under $B\to B+1$.  Consequently, both the partition
function of the generalized Coulomb gas and the charge-density
correlation functions remain invariant under the shifts $A+iB\to (A+iB)+in$,
for integer $n$, so that
\eqn\ZCGshift{Z_{CG}[A+iB+in, \zeta]=Z_{CG}[A+iB,\zeta],}
and
\eqn\rhoshift{\langle\tilde\rho^\sigma(\omega)
	\tilde\rho^{\lambda}(-\omega)\rangle(A+iB+in, \zeta)
        =\langle\tilde\rho^\sigma(\omega)
	\tilde\rho^{\lambda}(-\omega)\rangle(A+iB, \zeta).}
If we let $1/z=A+iB$ and $1/\tilde z = A+iB+in$, then we can use
these equations and equations \ZZCG\ and \tpcg\ to solve for the
transformation of the partition function and two-point function of the
Hofstadter model when $z$ goes to $\tilde z= z/(1+inz)$.  We find that
the partition function remains the same and the coordinate two-point
function transforms as
\eqn\tpfsf{\eqalign{
\langle\tilde{ \vec X}(\omega)
	\tilde{ \vec X}(-\omega) \rangle(z,V_0) =& \cr
           \tilde D(\omega;z-{i\over n})+&
    \omega^2 \tilde D(\omega;1+inz)\,
    \langle \tilde{\vec X}(\omega)
	\tilde{\vec X}(-\omega) \rangle({z\over 1+inz},V_0) \,
	\tilde D(\omega;1+inz).}}
We believe this transformation to be exact because we have summed to
all orders in perturbation theory. There are some subtleties having to do
with regulation, but they should have no influence on the long-time behavior.
In the next section, we will describe a discretized version of the dissipative
Hofstadter model for which duality transformations like \tpfsf\ are exact
without any qualifications.

The second Coulomb gas is obtained by making an instanton expansion of the
path integral. Here we assume that if the
friction is small enough, the main contribution to the path integral
comes from the ``approximate'' solutions to the Euclidean classical equations
of motion in the absence of friction.
In these solutions the particle moves between adjacent
mimina of the potential along instanton paths, which correspond to the
particle tunneling from one minimum to another. These frictionless instantons
for the Hofstadter model were studied in some detail in \defjah\ and we will
draw heavily on results presented there. When $M=0$, the classical
action for the instanton is given by $s = e^{-{|\beta|\over 2\pi}\gamma}$
with $\gamma\approx 7.33$, and when $\beta\sim 0$, it is
$s=4\sqrt{V_0Ma^2}/\pi\hbar$. When $M=0$, the instanton solution starting at
$(x,y) = (0,0)$ and ending at $2\pi\vec e_j$, with $\vec e_j =(\pm1,0)$ or
$(0,\pm1)$, is given by \defjah
\eqn\instsol{\vec f(\tau) = \vec e_j h(\tau)
                           +(\hat z\times \vec e_j)g(\tau),}
where
\eqn\htaudef{h(\tau)
    =2\tan^{-1}\left[\sqrt{2}\sinh(4\pi^2{V_0\over \alpha h} \tau)\right]+\pi,}
and
\eqn\gtaudef{g(\tau)
      =-i\cosh^{-1}\left[1+2\sech(8\pi^2{V_0\over \alpha h}\tau)\right].}
We can add together a series of these ``kinks'' to construct the
approximate solution
\eqn\appsol{\vec X_n(\tau) = \sum_{j=1}^n \vec e_j h(\tau-\tau_j)
                             +(\hat z \times\vec e_j)g(\tau-\tau_j),}
whose Fourier transform can be written in terms of the charge distribution
defined in equation \rhodef:
\eqn\qft{\vec X_n(\omega)
   = h_\omega\vec \rho_\omega + g_\omega(\hat z \times \vec \rho_\omega).}
When we substitute this solution back into the full action and then sum over
all such paths, we once again obtain a Coulomb gas expression for the partition
function: the two signs of the charge correspond to instantons and
anti-instantons; the locations of the charges correspond to the instanton
center of mass collective coordinate; the fugacity is given by the exponential
of minus the instanton action times a fluctuation determinant ($Ke^{-s}$),
where $K=V_0/\sqrt{|\beta|}$ when $\alpha=0$ and $\beta >> 1$; and the
instantons acquire long-range Coulomb interactions and phases from the
dissipation and magnetic field terms in the action.
The expression for the generating functional has the same structure as
before.  It is
\eqn\ZZgeninst{Z[z,V_0,F]=Z_{\rm gen}[S_{\rm inst}(z,F); Ke^{-s}],}
where now the energy function, $S_{\rm gen}$, in equation \ZZgen\ is replaced
by
\eqn\Sinstgen{S_{\rm inst}(z,F) =
      {1\over2}\sum_{jk}\vec e_j^\dagger \Delta(\tau_j-\tau_k;z)\vec e_k
         + \int{d\omega\over2\pi}\vec F_{-\omega}\cdot \left[
    h_\omega\vec \rho_\omega + g_\omega(\hat z \times \vec \rho_\omega)
              \right].}
Here, $\Delta$ specifies the interaction energy and phases between two
instantons and it is similar in structure to the Gaussian interaction function
$D(\tau_j-\tau_k;1/z)$.  In particular, its off-diagonal terms are exactly
the same, but the diagonal terms depend in detail on the shape of the
instanton solution. However, in the long-time limit
$|(\tau_j-\tau_k)4\pi^2V_0/(\alpha h)|>>1$, $\Delta$ turns out to
have a universal form
\eqn\deltdef{\Delta(\tau_j-\tau_k;z) \sim D(\tau_j-\tau_k;{1/ z}).}
Thus, in this limit the instanton interactions are identical to those
of a Coulomb gas with interaction parameter $z$ instead of $1/z$.
We will assume that properties of the critical points of the theory
are sensitive only to this long-time limit.  In that case, the
partition function can again be written in terms of the Coulomb gas
partition function given by equations \CGpfn\ and \fullSCG.  It is
\eqn\ZZCGinst{Z(z,V_0)=N Z_{CG}(z, K(V_0)e^{-s}),}
where $N$ denotes the value of the partition function in the absence of
tunneling.

We can again derive an expression for the two-point function in terms of
the charge correlation function of the new Coulomb gas:
\eqn\tptinstex{\langle\tilde X^\mu(\omega)
	\tilde X^\nu(-\omega)\rangle(z,V_0) =
      H^{\mu\sigma}(\omega)\,
   \langle\tilde\rho^\sigma(\omega)
	\tilde\rho^{\lambda}(-\omega)\rangle(z, Ke^{-s}) \,
      H^{\lambda\nu}(\omega),}
where $H(\omega)$ is
\eqn\Homegadef{H^{\mu\nu}(\omega)
         =h(\omega)\delta^{\mu\nu}+g(\omega)\epsilon^{\mu\nu}.}
The small $\omega$ (large $\tau$) behavior of $h$ and $g$ is such that
\tptinstex\ reduces, as $\omega\to 0$, to
\eqn\tptinst{\langle \tilde X^\mu(\omega)
	\tilde X^\nu(-\omega)\rangle(z,V_0) =
      \tilde D^{\mu\sigma}(\omega;1)\,
   \langle\tilde\rho^\sigma(\omega)
	\tilde\rho^{\lambda}(-\omega)\rangle(z, Ke^{-s}) \,
      \tilde D^{\lambda\nu}(\omega;1).}
The instanton method should be valid whenever the classical action
of the Hofstadter model (without dissipation) is large and the coefficient of
friction is much smaller than the classical action. We also assume that the
the large-time behavior of the system is not affected by the particular
(regulator-dependent) small-distance form of the propagators.

To summarize, the preceding calculations demonstrate that the original
partition function is equivalent to the partition function for
two different Coulomb gases of ``$X$'' and ``$Y$'' particles
which correspond to the $X$ and $Y$ components of $\vec e_j$, respectively.
The particles have charges $\pm 1$ and a fugacity which depends on which
Coulomb
gas is under consideration. The structure of the interaction energy is the
same for both gases: like particles (both $X$ or both $Y$) interact via
$-2Ae_i^\mu e_j^\mu\log|\tau_i-\tau_j|$ for $|\tau_i-\tau_j|>>1$, and unlike
particles via $i\pi B\sign(\tau_i-\tau_j)\epsilon^{\mu\nu}e_i^\mu e_j^\nu$.
When we treat the potential as a perturbation, the fugacity is given by $V_0/2$
and the interaction strengths by $A+iB = 1/z$ (where $z$ is the parameter
introduced in \twopt ). When we use the instanton
method, the system is described as a gas with fugacity $/zeta=K(V_0,z)e^{-s}$
and interaction strength $A+iB = z$.  Therefore, we have established that
the system with dissipation and flux per unit cell given by $z$ and with a
potential of strength
$V_0$ is dual to the system with $z\to1/z$ and a new potential
$\tilde V_0/2=K(V_0,z)e^{-s}$.

The two-point function of the original system can be expressed in terms of the
charge-density correlation function for either of the two Coulomb gases
described above. Upon eliminating the charge density correlators between
the two expressions, we get
the following duality relation between two-point functions of the original
theory in dual regions of parameter space (we expect it to be valid for large
times, or, equivalently, small $\omega$):
\eqn\tpinv{\langle \tilde{\vec X}(\omega)
	\tilde {\vec X}(-\omega)\rangle(z,V_0) =
           \tilde D(\omega;z)-
    \omega^2 \tilde D(\omega;z)\,
    \langle \tilde{\vec X}(\omega)
	\tilde{\vec X}(-\omega)\rangle({1/ z},\tilde V_0)\,
	\tilde D(\omega;z)~.}
Note that the fugacity transforms along with $z$ ($V_0\to\tilde V_0$).
This sort of relation has the strongest consequences in circumstances
where there is in fact no dependence of the two-point function on fugacity.
As we shall see, this covers a lot of territory.

Lastly, we may combine the ``exact'' duality transformation $z\to z/(1+inz)$
with the approximate duality transformation $z\to 1/z$.
A complication here is that under the
second duality, $V_0$ transforms in a $z$-dependent way.
Consequently we learn  that the dissipative Hofstadter system
at $z$ and $V_0$ is dual to the system at $z + ni$ and some other potential
strength. The two systems should have the same partition function up
to an analytic prefactor.
Additionally, if for some reason  the two-point function is not explicitly
dependent on $V$, then, by composing \tpinv\ and \tpfsf, we obtain
\eqn\tpshift{\langle \tilde{\vec X}(\omega)
	\tilde{ \vec X}(-\omega) \rangle(z+in) =
            \langle \tilde{\vec X}(\omega)
		\tilde{\vec X}(-\omega) \rangle(z)~.}

We conclude that the system has an exact symmetry under $z \to z/(1+inz)$ and
an approximate symmetry under all compositions of $z \to 1/z$ and $z \to z+i$.
We will draw the consequences of this for the phase structure of the theory
in the final section.

\newsec{The Discrete Gaussian Model: an Instructive Example}

Because our model has an approximate duality under $z \to z+i$ and
$z \to 1/z$, it is natural to suspect that there is a similar model which
has an exact duality under these transformations.  The discrete gaussian
chain with the interaction $V(r) = \alpha/(r^2+1/4)$ is known to be exactly
dual to one with $\alpha$ replaced by $(1/\alpha)$ \kjaer\ .
We find that if we
couple two such chains via a magnetic interaction, the system is then exactly
dual under $z \to z+i$ and $z \to 1/z$.  We define such a system as follows.
There is a one-dimensional lattice located at the sites $i=1,2,...,N$ and
a second dual lattice at $i' = 1/2, 3/2,...,N-1/2$.  At each site there are
height variables $h_i^x$ and $h_{i'}^y$ with $h_{i+N}^x = h_i^x$ and
$h_{i'+N}^y = h_{i'}^y$.  The height variables can take on any integer value
except for $h_N^x$ and $h_{N-1/2}^y$, which we set to zero in order to
keep the partition function finite.  We will define our discrete Gaussian
Hamiltonian with magnetic interaction to be
\eqn\DGH{\eqalign{H_N^{DG}=&{1\over 2} \alpha\sum_{i\ne
j}V_N(i-j)(h_i^x-h_j^x)^2
            +{1\over 2} \alpha\sum_{i'\ne j'}V_N(i'-j')(h_{i'}^y-h_{j'}^y)^2\cr
   &-{1\over 2} 2\pi i\beta\sum_{i}h_i^x(h_{i+{1\over2}}^y -h_{i-{1\over2}}^y)
+{1\over 2} 2\pi i\beta\sum_{i'}h_{i'}^y(h_{i'+{1\over2}}^x
-h_{i'-{1\over2}}^x)
   ,}}
where $V$ is chosen so that its Fourier transform $\tilde V$ satisfies
\eqn\Wdef{W_N(k)=\tilde V_N(0)-\hat V_N(k) = 2\pi |\sin{k\over2}|.}
The partition function
\eqn\ZDGdef{Z_N^{DG}[z] = \sum_{h_i^x}\sum_{h_{i'}^y}e^{- H_N^{DG}},}
depends on two parameters $\alpha$ and $\beta$, which we have combined
into a complex parameter $z=\alpha+i\beta$. With these definitions, the
momentum space version of \DGH\ reads
\eqn\ftH{H_N^{DG}={1\over2}\sum_k
          \vec h_{-k}\cdot \tilde D^{-1}(k;z)\cdot \vec h_k}
where $h_k^x$ and $h_k^y$ are the Fourier transforms of $h_j^x$ and $h_{j'}^y$,
\eqn\hxft{h_j^x= \sum_k e^{i k j} h_k^x,~~h_{j'}^y = \sum_k e^{i k j'} h_k^y,}
(where $k=2\pi r/N$ and $r$ runs over the integers satisfying $-N/2<j\le N/2$),
and $\tilde D^{-1}(k;z) = 2 W_N(k)M(k;z)$, with $M$ defined by
\eqn\Mdef{M(k;z) = \pmatrix{\alpha&\beta {k\over|k|}\cr
                            -\beta {k\over|k|}&\alpha\cr}.}
$\tilde D(k;z)$ is the discrete-time version of the propagator
$\tilde D(\omega;z)$ originally introduced in \twopt .

Now we can do a Poisson resummation trick, exactly as in \kjaer, to reexpress
the discrete Gaussian system as a Coulomb gas. The procedure is to
replace the $h_i$ and $h_{i'}$ by continuous variables
$\nu_i$ and $\nu_{i'}$ (over which we will integrate) but to include delta
functions of the form $\sum_{q_j=-\infty}^\infty e^{2\pi i q_j\nu_j}$ to
pick out the integer values of $\nu$.  On integrating out the $\nu_i$'s
we obtain
\eqn\ZDGCG{Z_N^{DG}[z]=NC_N(\alpha^2+\beta^2)^{-(N-1)/2}
                            Z_N^{CG}[{1/ z}],}
where $C_N = \prod_{k\ne0}\pi/W(k)$ and $Z_N^{CG}$ is given by
\eqn\ZNCGdef{Z_N^{CG}[{1/ z}] =
               \sum_{q_i^x}\sum_{q_{i'}^y}
  \exp\left[-\sum_{k\ne0}{\pi^2\over W_N(k)} \vec q_{-k}\cdot
                 M(k;{1/ z})\cdot\vec q_k \right].}
Here, $\vec q_k=(q_k^x, q_k^y)$, where $q_k^x$ and $q_k^y$ are the
Fourier transforms of $q_i^x$ and $q_{i'}^y$, respectively.
The $q_i$ are constrained to satisfy $\sum q_i^x = \sum q_{i'}^y = 0$.
It is not too hard to recognize \ZNCGdef\ as a discretized generalized
Coulomb gas. Indeed, in the limit $N\to\infty$, the interactions between
the charges are exactly the same as in the model of the previous section.
The key differences are that: the charges can now take on any integer value
(they were previously restricted to $\pm 1$); there is only one charge per
lattice site; and the fugacity is $1$ rather than being freely variable.

Next, we perform Cardy's transformation \cardy\ on the expression for
$Z_N^{CG}$. We define the new height variables, $l_{i'}^x$ and
$l_i^y$, by
\eqn\ldef{q_i^x=l_{i+1/2}^x-l_{i-1/2}^x,
      \quad q_{i'}^y =l_{i'+1/2}^y-l_{i'-1/2}^y,
       \quad {\rm and} \quad l_0^y = l_{1/2}^x = 0.}
On making this variable substitution in \ZNCGdef\ we find,
on comparing with \ftH ~,
that we have reproduced $Z_N^{DG}$ with $z\to 1/z$. Specifically,
\eqn\ZCGdual{Z_N^{DG}[z] = NC_N(\alpha^2+\beta^2)^{-(N-1)/2}
                               Z_N^{DG}[{1/ z}].}
This exact duality between the system at $z$ and $1/z$ depends on the precise
form \Wdef\ and \Mdef\ of the interaction. The partition function in equation
\ZDGdef\ is
also manifestly invariant under $z\to z+i$. Thus, the system is exactly dual
to itself under all compositions of the two transformations
$z\to1/z$ and $z \to z+i$. We will shortly explore the consequences of this.

By a slight variant of the above arguments, one can derive, following \kjaer~,
an exact duality relation for the two-point function:
\eqn\DGtp{\langle \tilde h(k)^{\mu} \tilde h(-k)^\nu\rangle(z) =
          \tilde D^{\mu\nu}(k;z)-
       M^{\mu\rho}(k;{1/ z})\,
      \langle \tilde h(k)^\rho \tilde h(-k)^\lambda\rangle({1/ z})
     \,  M^{\lambda\nu}(k;{1/ z}).}
The duality relation \tpinv\ for the continuum problem is equivalent to this
if, for some reason, the two-point functions don't depend on the potential
amplitude or there is some self-dual value of the fugacity for which
$\tilde V=V$.

At the self-dual point, $z=1$, equation \DGtp\ determines the two point
function exactly. It is
$\langle\tilde h(k)^\mu \tilde h(-k)^\nu\rangle(1)={1\over 2}\tilde D(k;1)$,
which is half the value for the continuous gaussian model. (Note that there
is a certain discontinuity here: anywhere inside the first circle, no matter
how close to $z=1$, the two-point function is given by $\tilde D(k,z)$.) We
can apply the transformations $z \to z+i$ and $z \to1/z$ to this expression
to find the value of the mobility for all $z$ given by
\eqn\ztan{z = {d-ib\over a+ic}= {1\over a^2 + c^2}-i{ab+cd\over a^2+c^2},
   \quad {\rm for} \quad ad-bc=1, \quad a,b,c,d\in Z.}
It is
\eqn\tptan{\langle h_k^\mu h_{-k}^\nu\rangle(z)={1\over 2}(a^2+c^2)
	\tilde D(k,1).}
We observe that at these special points (which include the special points
on the first circle mentioned in the previous section) the off-diagonal
propagator (or the Hall coefficient) vanishes even though the magnetic field
at these points is not in general zero.

\newsec{Duality and the Phase Diagram}

Now we want to use these results to map out the phase structure of the
dissipative Hofstadter model in the $(\alpha,\beta)$ plane. Along the
zero-field line, there is no doubt that the system is localized for
$\alpha > 1$ and delocalized for $\alpha < 1$. The Coulomb gas
treatment of Section~4 allows us to expand on this somewhat. There are
two `dual' Coulomb gases: in the first, the charges correspond to
insertions of the cosine potential and the fugacity is one-half of
the potential strength
$V_0$; in the second, the charges correspond to instantons describing
tunnelings between potential minima and the fugacity is essentially the
hopping probability of the particle. The one-loop renormalization group
argument applied to a Coulomb gas says that fugacity scales to zero (and
the associated charges become irrelevant) when the coefficient in front of
the logarithmic self-interaction energy is greater than a critical value.
Applied to the first Coulomb gas, this argument says that the periodic
potential insertions are irrelevant everywhere within the circle
$\alpha^2+\beta^2=\alpha$. Applied to the second Coulomb gas, it tells us
that the fugacity for instantons scales to zero (i.e. the hopping operator
is irrelevant and the particle is localized) everywhere to the right of
$\alpha=1$. Thus, it seems very plausible that the system is delocalized
within the `first circle'  $\alpha^2+\beta^2=\alpha$ and localized in the
half-plane to the right of $\alpha=1$. As partial confirmation of this, recall
that, in Section~3, we were able to exactly calculate
the two point function at those points on the `first circle' where
$\beta/\alpha \in Z $ and indeed found delocalized behavior.

With the help of duality arguments, we can parlay these two results into a
global picture of the phase structure. There are two levels of duality
which we could imagine using: The discrete Gaussian model has an exact
invariance under the two-element modular group $z\to z+i$ and $z\to 1/z$,
as described in Section~5. The Hofstadter model, however, is exactly
invariant only under the subgroup $z\to z/(1+inz)$ (for $n$ integer),
as described in
Section~4. What's missing for the Hofstadter model are the
transformations $z\to z+ni$ or $z\to 1/z$: under those parameter
transformations, the strength of the
periodic potential transforms as well. However, in the regions where the
potential is effectively irrelevant (not just inside the first circle, as
we shall see), the shifts and inversions should still be valid
symmetries of the asymptotic
behavior of the theory.

As a first step, let us see how far we can get with the exact symmetry of the
Hofstadter model. Under $z\to z/(1+inz)$, the half-plane $\alpha >1$ is mapped
into the circle of radius $1/(2n^2)$ and center $1/(2n^2)-i/n$. This is
tangent to the first
circle at a point where $\beta/\alpha=-n$ and also to the $\beta$-axis at
$\beta=-1/n$. These circles are displayed in Fig.~1. Now we use the exact
relation \tpfsf\ to compute the mobility (i.e. the long-time or low-frequency
limit of the two-point function) in the circles in terms
of the mobility in the region $\alpha >1$. Since this latter is a region of
localization, its mobility is zero, and \tpfsf\ expresses the
mobility in the circles in terms of the mobility of the Gaussian theory,
encoded in \twopt~. The result is
\eqn\nthmob{\langle X^\mu(t)X^\nu(0)\rangle(z,V)= D^{\mu\nu}(t;z+{i\over n})
       \qquad {\rm for} \quad |t|>>0,}
which is to say that, as far as mobility is concerned, the n-th circle is
just a translation parallel to the $\beta$ axis of the Gaussian mobility
evaluated within an appropriate sub-circle of the first circle. The simple
renormalization group argument implies that the mobility is indeed Gaussian
within the first circle, so this is consistent with the $z\to z+i$ duality.
These results are completely different from what one would
find if the periodic potential \perpot\ were zero: in that case the mobility
would be given exactly by \twopt\ everywhere in the $z$-plane. So, although
there is no explicit dependence on $V_0$, the potential is clearly relevant
outside the first circle!

Because the mobility inside the first circle and all the circles shown in
Fig.~1 has no explicit $V$-dependence, we should be able to
use the duality relation under $z\to z+in$ (equation \tpshift)
to find the mobility
for all points that are translates of these regions.  As indicated
above, this symmetry is probably correct for the asymptotic behavior
of the Hofstadter model.  This claim is supported by the observation
that for the two circles immediately above and below the first circle,
we have just shown that the mobility is $D(t;z\mp i)$, which is exactly
what we would expect from the duality transformations given by
equation \tpshift.  When we apply the translation by $im$ to the first
circle, we obtain a chain of circles tangent to each other and to the
lines $\alpha=0$ and $\alpha=1$.  Next, we can repeatedly apply
$z\to z/(1+inz)$ and $z\to z+in$ to these circles, which starts filling in
the regions between the line $\alpha=0$ and the first chain of circles
with new circles that are tangent to the $\beta$-axis and already
existing circles.  The result is shown in fig.~2.
It is the same phase diagram found by Cardy for abelian lattice gauge
theories with theta terms \thetar.  In the diagram,
between any two circles and the $\beta$-axis,
there is another, smaller circle.  So, the boundaries of these circles,
which, as explained below, should identify lines of phase transitions,
get denser and denser as one gets closer and closer to the $\alpha=0$
line (the zero dissipation line).  On this line, every point with
rational $\beta$ is tangent to a circle, and every point with
irrational $\beta$ is not.  This connects nicely to the
Hofstadter model (with no dissipation) where the behavior is quite
different depending on whether $\beta$ is rational or irrational.

According to the equations \tpfsf\ and \tpshift\ for the transformation
of the two-point function, we find that inside all the circles the
mobility is Gaussian: in a circle tangent to the $\beta$-axis at
$z=p/q$, with $p$ and $q$ relatively prime, the mobility is given by
$D(t;z-ip/q)$.  The triangular regions between the circles belong
to some other phase which this argument doesn't help us identify.

Along the phase transition circles, one expects to find a non-trivial
critical theory. Each of these circles is a mapping of the critical
line at $\alpha=1$ (or the critical circle $\alpha^2+\beta^2=\alpha$)
by compositions
of the transformations $z\to z+i$ and $z\to1/z$.  Under
these transformations, the partition function is multiplied by an
analytic function (which depends on $C$ and $N$ in equations \ZZCG\
and \ZZCGinst)
and the strength of the periodic potential changes.  Because the
existence of the phase transition on the first circle is independent
of $V_0$ (at least for small $V_0$), neither of these two changes to the
partition function should affect the fact that it is critical.
Consequently, we expect all the circles to be critical circles, at
least for some range of values of the potential strength.  However,
because we do not yet know exactly how $V_0$ changes under these
transformations, we know little about the critical theories, except
at the points where $\beta/\alpha =n$ for $n$ an integer,
which is where the other circles
are tangent to the first circle.  For these points, we can apply the
exact duality transformation \tpfsf\ to the two-point function
at $z=1$ (given in equation \toopt) and then simplify the resulting
expression by using equations \Madd, \Mmult, and \Mcmult to obtain
\eqn\sptpt{{\langle\vec X(t)\vec X(0)\rangle}(z_n,V_0)=
       D(t;z_n)+\left(f(V_0)-1\right)D(t;z_n^2)}
where $z_n=1/(1+in)$ and $f(V_0)$ is the factor by which the two-point
function at $z=1$ differs from the Gaussian two-point function \twopt .
With the specific regulator used in Sect.~3, we found that
\eqn\fdef{f(V_0)=
    ( 1+ ({V_0T\over2}e^{-\langle X(0) X(0)\rangle})^2)^{-1}=
	(1+(\tilde V_0/2)^2)^{-1}~,}
but, in general, its exact form depends on how the theory is regulated.

We note that this two-point function depends on the strength of the
potential.  For the discrete Gaussian model, there is no parameter
corresponding to $V_0$, but
the two-point function is consistent with equation \sptpt\ with
$f(V_0)= 1/2$ or $\tilde V_0= 2$.  For this special value of $f(V_0)$, the
two-point function at $z_n = 1/(1+in)$ becomes
\eqn\xstp{{\langle\vec X(t)\vec X(0)\rangle}(z_n,V_0)={1\over2}(1+n^2)D(t;1).}
A point to emphasize here is that the off-diagonal
part of the mobility (the Hall coefficient) vanishes at these points
for this particular value of $V_0$, even though the magnetic field is not zero.
These considerations make it quite plausible that the discrete Gaussian
model of Sect.~5 has the same critical behavior as the continuum dissipative
Hofstadter model for a special value of the potential strength.

In sum, the potential strength is an
exactly marginal variable. The higher-point functions for $\alpha=1$
reduce to products of
two-point functions plus contact terms (which have not been explicitly
computed) and for the other points on the first circle we have evidence
that some of the higher $n$-point functions are non-zero, SU(1,1) invariant
functions.  Therefore, even if we don't know the details, we are
guaranteed that the critical theory is not identical to the Gaussian theory

\newsec{Loose Ends and Problems for the Future}

The main result is that we have discovered a phase structure for the
Hofstadter model (and a discrete version of it) which shows how the fractal
energy level structure of the pure quantum mechanics is matched by a fractal
structure of phase transitions in the dissipative theory.
There are several issues to pursue. We have identified critical surfaces
with critical behavior depending on at least two parameters: potential
strength and arc position in the $\alpha$-$\beta$ plane.
We know little about the detailed
properties of these critical theories and it would be very interesting
to study them in more detail, especially in the discrete Gaussian model,
where things should be simple. We have seen in Section~3
that at the special points
on the first circle something remarkable happens. We have hints that at
these values of $\alpha$ and $\beta$, most of the $n$-pt functions (for $n>2$)
reduce to contact terms, which is reminiscent of topological field theory,
but we do not yet know how to pursue this connection. These
critical theories have an interpretation as solutions of open string theory
with nontrivial backgrounds. It will obviously be interesting to explore
the spacetime interpretation of these states. We also want to understand
the nature of the phases in Fig.~2 which fill in the interstitial
regions between the circles. They
will presumably be some sort of Coulomb phase which is neither localized
nor delocalized. We have yet to determine what this means quantitatively
for the behavior of the mobility and Hall coefficient.

Another line of development takes us into N-body physics. We have seen
that the transport coefficients (especially the Hall coefficient) of a single
electron
have remarkable structure in the presence of dissipation. We can actually
turn the Hall effect off by choosing the right potential amplitude at special
values of the field. Filling Landau levels with such electrons would doubtless
lead to new quantized Hall effect phenomenologies. There is a speculative
connection with nonperturbative open string theory, because inserting many
boundaries into the string possibly corresponds to having many electrons
in the dissipative quantum system.

The final big issue is the possibility of
constructing experimental realizations of the physics we have uncovered.
Work is already under way to observe, separately, the Hofstadter spectrum and
the macroscopic tunneling effects for a two level system.  It may be
possible that specially constructed junction arrays could be
used to experimentally investigate the combined effects of the magnetic
field, periodic potential and dissipation.

\bigbreak\bigskip\centerline{{\bf Acknowledgements}}\nobreak
This work was supported in part by DOE grant DE-AC02-84-1553.
D.F. was also supported by the Department of Education and DOE grant
DE-AC02-76ER0369.

\vfill\eject

\listrefs

\bye